# Who Will Win Practical Artificial Intelligence? AI Engineerings in China

**Authors:** Huai-Yu Wu[1], Feiyue Wang[2], Chunhong Pan[1]

**Affiliations:**

[1] National Laboratory of Pattern Recognition, Institute of Automation, Chinese Academy of Sciences, Beijing, 100190, P.R. China.

2 The State Key Laboratory of Management and Control for Complex Systems, Institute of Automation, Chinese Academy of Sciences, Beijing, 100190, P.R. China.

**Running Head:** AI has become one of the most popular focus in China.

**Abstract**: Currently, Artificial Intelligence (AI) has won unprecedented attention and is becoming the increasingly popular focus in China. This change can be judged by the impressive record of academic publications, the amount of state-level investment and the presence of nation-wide participation and devotion. In this paper, we place emphasis on discussing the progress of artificial intelligence engineerings in China. We first introduce the focus on AI in Chinese academia, including the supercomputing brain system, Cambrian Period supercomputer of neural networks, and biometric systems. Then, the development of AI in industrial circles and the latest layout of AI products in companies, such as Baidu, Tencent, and Alibaba, are introduced. Last, we bring in the opinions and arguments of the main intelligentsia of China about the future development of AI, including how to examine the relationship between humanity on one side and science and technology on the other.

**Main Text:**

Although there were ups and downs for AI throughout its history[1], in recent years, many breakthroughs in AI research have been achieved worldwide, from the Watson system of IBM, to Siri of Apple Inc., and to Google Brain. In 2016, an AI program named AlphaGo[2] had defeated professional human champion of Chinese Go which was developed by Deepmind, a Google AI team. Such achievements have primarily displayed a brain-like intelligence trend with a focus on self-learning.

There are several main factors driving AI field to current level. Firstly, thanks to the great decreases of computing cost. Inexpensive parallel computing allows the connection numbers of neural network nodes to be of hundreds of millions. Secondly, it's because of the emergence of big data. Every kind of AI needs enough samples for education. As estimated, the total data amount generated in 2015 worldwide is more than 20 times than it was 10 years ago. Such great amount of data will provide enough materials for machine to learn. Another more important reason is the progress of the algorithms. Currently, the most advanced technologies in AI field are Deep Neural Networks (Deep Learning). The deep learning algorithm models also have went through a fast iteration period: Deep Belief Network, Sparse Coding, Recursive Neural Network, Convolutional Neural Network (CNN), and more.

For the moment, those kinds of companies focusing on specific application for deep learning or machine learning are the largest in quantity, then the following are those kinds of natural language processing companies which focus on voice recognition, as well as computer vision companies. Major big companies like Facebook, Google, Yahoo and Baidu are all trying to apply deep learning algorithms into their product development, in the purpose of making the product more intelligent, and enhancing user experience. Participators who are willing to work in AI field must think carefully about two factors, which are also the main basis for this article to select those AI projects. Firstly, what does AI change? Great products or technologies must have the ability to shift the customers' behaviors in some circumstances, otherwise a product does not provide enough reason for customers to choose if it does not matter too much to own it or not. If a product's core competitive force is AI, it should match perfectly with the scenario. The core of AI related products lies in to how much level it can substitute for mankind functionalities or enhance working efficiency. For example, a radiologist can get a job offer of 300 thousand US dollars annually in the United States after receiving 13 years' education and training. While the latest pattern recognition software can finish the majority of the work but with a cost of only 1/100. This kind of example also applies to layers, bank employees, website editors and drivers, to name a few.

Secondly, whether the AI technology in the chosen direction is practically possible. Many AI technologies have not been very mature yet, it's impractical to try to solve all the problems only by pursuing technology breakthroughs. Instead, we could start by working on some specific scenarios if it is not working in a generic scenario. We could start by doing a product which is possible if a universal product is not possible yet to have.

AI contains many advanced technologies and branches; it is hard to define it within one sentence. And it is developing all the time, for example some so-called AI stuffs in the old times now are just simple mechanical repetitions from current perspective, like automatic washing machine and automatic door, etc. Actually, for the moment the effects caused by AI revolution to white-collars, like website editors, are just like the effects caused by machine technologies to blur-collars, like textile workers in the old times. Generally speaking, the nature and the goal of AI do not change too much all the time, which is to help mankind to finish partially intellectual work/task. In this sense, AI is more like an engineering issue, not simply a scientific issue. It is very important to understand this point, which is also the reason why this article brings the view into China. As we all know, China is still not in the same level in science field comparing with western countries, but in engineering field it is starting to rank tops in the world and is becoming the world biggest factory, for example in the aspects of high speed train, aerospace and mobile internet. In fact, speaking from the national level, AI is a very suitable giant project to a country, especially in those aspects like high-performance computing, big data, labor cost, which is good to receive national resources for supporting, so from this point of view, it is particularly suitable for China.

### Artificial Intelligence Engineerings in China

As a developing country, China is still a follower in the field of AI science, but it does not mean China is behind western countries in the field of AI engineerings. In fact, China has a good faith in Pragmatism, who is good at utilizing western countries' AI science into a very deep level on engineerings. Although these kinds of AI projects own seemly ordinary theoretical frameworks, but the success or failure of an AI project usually lies in the details of engineering's (such as the layers of neural networks, the selection of the parameters, and the combinative mode of varied methods). Chinese people are really good at building a new hybrid engineering framework after a full analysis of different theoretical frameworks' merits and faults, in order to have a better performance.  An interesting case, there is a Hollywood movie named *2012*, in which people need to build giant Noah's arks to escape from the doomsday. These arks are finally built by Chinese people, because only Chinese people can make it possible to finish such a huge and complicated project within a short time.

In recent years, in the development of its economy, China keeps investing more into the field of scientific technology to achieve industrial transformation, i.e., from "Made in China" to "Intellectually Made in China". Following the introduction of big data, cloud computing, Internet and mobile networks, AI has become the new flash point in China. Chinese President Xi Jinping emphasized that the robot revolution will be an important starting point of the Third Industrial Revolution[3]. Premier Li Keqiang further proposed the concept of Internet + in his government work report[4]. And the Chinese government has formally approved the "China Brain" plan and regards it as one of the major projects concerning the future of Chinese development.

"Whoever wins AI, will own the future", said Dr. Andrew Ng, the leader of the original "Google Brain" project, now the leader of the "Baidu Brain" project. In this paper, we will emphasize China's focus on AI and some arguments in the industry.

**Supercomputing Brain System**

The Institute of Automation, Chinese Academy of Sciences (CASIA), which was established in Beijing in October 1956, is the top research institution in China in the field of intelligent information processing. CASIA founded the "General Department of Supercomputing Brain System" in 2012 and further founded "Research Center for Brain-inspired Intelligence" in 2015, triggering research on human-like brain engineering. This research mainly focuses on three areas: the atlas of the brain network, brain simulation, and brain-like computing systems. This department has been engaged in cooperation with Tsinghua University, Xi'an Jiao Tong University (XJTU) and several other top universities in China to jointly persuade the government to launch the "China Brain" project, which will be a research plan that receives financial aid, as much as 1 billion USD.

"In the aspect of multi-dimensional simulation[6], especially the multi-encephalic region and multi-channel coordination simulation, our research work should be distinctive. And I hope in the near future it would become the representative technology for China, able to compete with brain simulators from Canada and EU," said Prof. Zeng Yi, the leader of the team. In practice, the team also paid much attention to international cooperation with related units of other international brain plans. For example, the deep cooperative efforts with Professor Wang Yun from the EU Human Brain Project (HBP) team, especially the cooperative research results in the field of morphology repair of nerve fibers, were not only jointly published in the literature but were also prepared to be included in an application for the HBP project.

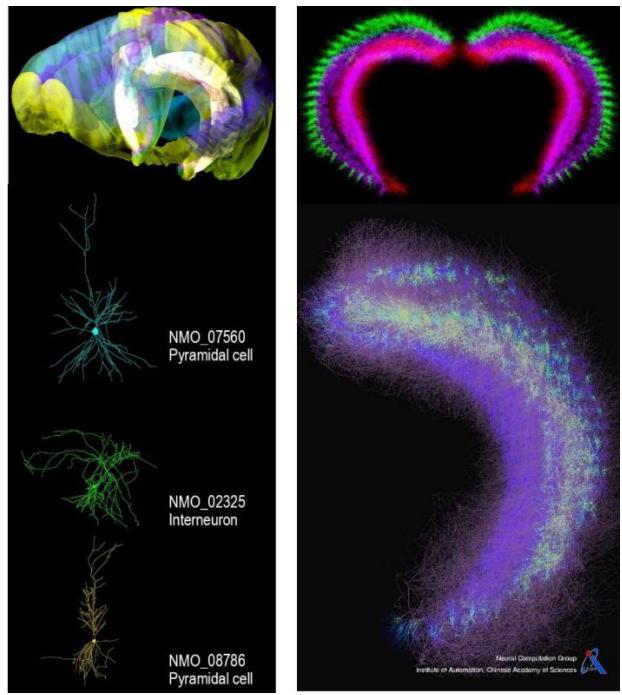

**Fig. 1**. Brain simulation and calculation: composition of each encephalic region of the mouse brain (left); morphology of the hippocampus of the mouse brain (upper right) when the hippocampus of the mouse brain implements the task of memory; the condition in which the neuron network gives off electricity is shown in the picture (lower right).

At the same time, the Brainnetome Center is a brain-net research team that aims to analyze the fine divisions, connecting modes and functions of different encephalic regions, particularly the dynamic evolution rules of the brain network. "The brain function is not completed independently by a single neuron or just one encephalic region, but realized thought the interactive function of the neuron body (group) within the neural circuit, functional column or encephalic region," emphasized Prof. Jiang Tianzi, the director of the center. In 2010, his team proposed the brain-connecting group / net group plan[10] (corresponding to the plan of the gene group in 1990, and the plan of the protein group in 2001) for the first time in the world. Now, they have finished the new atlas of the brain, the Brainnetome Atlas, which includes the division into sub-regions of the encephalic region functions in the dimension of the whole brain and in terms of multi-mode connectivity[5].

### Cambrian Period Supercomputer of Neural Networks

To achieve "ubiquitous" AI, besides an effective algorithm, the support of portable hardware with high performance and low energy consumption is needed as well. The Future Computing Team of the State Key Lab of Computer Architecture, which is affiliated with Institute of Computing Technology, Chinese Academy of Sciences, led by Professor Chen Yunji, optimizes the structure of processors by making use of machine learning and evolutionary algorithms; this team has developed a type of machine learning and arithmetic device by adopting the artificial architecture of a neural network – a Cambrian Period neural network computer. This computer can process a neural network of any scale and depth at high efficiency and has reached a processing speed for the neural network that is as fast as 100 times the previous registered speed in only less than 1/10 of the space and power consumption of the traditional processors. Both the performance and power ratio were enhanced 1,000 times compared with earlier technology. The Cambrian Period No.1 (English name: DianNao) was awarded the Best Paper Prize of ASPLOS 2014[7] – the top conference in the field of computer architecture.

Furthermore, the deep learning and processing speed of the Cambrian Period No.2 processor is 21.38 times as fast as the mainstream GPU (NVidia K20M), becoming the quickest deep learning chip in the world with less than 1/330 the GPU power consumption[8].

Besides portable devices, it is worth to mention that according to the TOP500 ranking list of supercomputers[14], "Sunway TaihuLight" developed by China had reached the peak speed of 125,436 TFlop/s which was the supercomputer of the most powerful performance across the world. It is possible that such supercomputers can become applicable in future AI calculations on a very large scale.

### Biometric Systems

The vice President of the Chinese Academy of Sciences, Professor Tan Tieniu think that biometrics will be the key techniques in the intelligent time. In September 2012, *Science* magazine published *China's Sharp Focus On Biometrics*[9], which used pictures to explain the text on the contents page and briefed the reader on the development of China's biometric system with respect to the international frontier. This article used a large amount of space to introduce the research on video data analysis conducted by NLPR (National Laboratory of Pattern Recognition) and its application in public security. The core of the innovation is human face recognition by near-infrared and multispectral technology, which can be used to solve the difficult problems of varying illumination and anti-faking of the human face. In that volume of *Science* magazine, it concluded: "The leading Chinese image-processing and machine-learning labs now compete with the top institutions in the U.S.".

### Focus on AI in Industry

The area of "big data" plays an important role in improving the performance of machine learning and algorithms. In today's China, the three largest internet enterprises, Baidu, Alibaba and Tencent, are called BAT. Such internet enterprises were born with advantages in obtaining "big data".

Baidu invited Dr. Andrew Ng to join. Once an Associate Professor of Stanford University, he also led the "Google Brain" project and now works as the chief scientist leading the Baidu Research Institute. The reason that Andrew Ng joined Baidu was that he valued the ambition of Baidu toward developing deep learning, and deep learning was regarded as the best method to absorb and analyze "big data". Until now, there was no other company that could launch a product on deep learning faster than Baidu. The computing speed of the "Baidu Brain" will be 100 times as fast as the speed that "Google Brain" was in 2012. Baidu has established the largest deep neural network in the world, with 20 billion parameters, and adopted the method of deep learning in 2013, aiming to improve the ranking quality of webpage searching to the greatest extent. Andrew Ng hoped to utilize the big data of Baidu to overcome problems such as unsupervised recognition of massive objects and the understanding of natural languages. For example, Baidu launched a product to identify pictures, which was called "Baidu picture-reading". Users only needed to take a photo of any object (e.g., a handbag of an unknown brand or an unheard-of plant), and, using their mobile phone, Baidu would return detailed information on the item, where to buy it and the prices.

As a Chinese company listed on the NYSE, Alibaba had over 100 PB of data, including the users and commodities data, which is equivalent to 4 Seattle Central Libraries. Alibaba also actively applied machine learning methods to optimize the searching and advertisement system, CTR estimation, webpage ranking of the searching results and personalized system to enhance BI (Business Intelligence). Currently, Tencent, China's largest instant messaging software company, has released a platform for deep learning − Mariana. Tencent has over 848 million monthly active users; social data in such large amounts could be used to analyze and learn human emotions − joy, anger, sorrow and happiness, and could be developed to correspond to the human EQ.

In one of China's most famous IT companies, Huawei, the director of its Noah's Ark Laboratory, Yang Qiang, a Professor from the Hong Kong University of Science and Technology, unveiled the research achievement of the lab for the first time on October 2014 − an AI system called MoKA. Transferring the experience and knowledge of the past to learning in the present using a computer would become the "soul mate" of self-learning that could accompany humans all of their lives. In addition, another speech recognition company worth mentioning is iFLYTEK; the capacity for speech recognition in the Chinese language, as developed by this company, has surpassed major international companies, for example, that of Siri of Apple Inc.

It is worth to mention that Microsoft China located in Beijing also plays an important role in the development of Chinese AI Industry. In the 2015 ImageNet challenge, Microsoft Research Asia (MSRA) researchers announced a major advance in technology designed to identify the objects in a photograph or video, showcasing a system whose accuracy meets and sometimes exceeds human-level performance[12]. Their system was very effective because it allowed them to use extremely deep neural nets with 152 layers, which are as much as five times deeper than any previously used. Moreover, in 2014, Microsoft STC produced an AI chat robot, called Microsoft Xiaoice[13], who has accomplishing 600 million conversations with 37 million Chinese peoples via Internet until now.

### Exploitation of AI in China

The application threshold of AI technology is high. For example, the development of AI requires professional background knowledge and the ability to address the difficulty of performing calculations on large amounts of big data. Therefore, it is still difficult to obtain a broad application in the market, and many AI research projects are financially assisted by the government and are directed by the government. For the processing of natural language, state-run research institutes cooperate with the Sina Weibo and Tencent Wechat, which are the Chinese versions of Twitter and WhatsApp, respectively, conducting assessments of users' credit and performing analyses on public opinions in the context of social networking to regulate the environment of public opinion on a social network platform.

At the same time, many of the AI research projects could also be used for military purposes. Using the four-foot robot development as an example, the Center for Robotics, Shandong University and the National University of Defense Technology have both been close to reaching the technical level of the BigDog robot of the American military industry. Moreover, the research on a simulated robot-fish in China has been implemented over the past ten years[15]. The designed robot-fish can implement high-mobility controls, such as pitching, snorkeling, turn-back, quick-start, changing speed and direction during movement, inverse swimming, depth-keeping and braking. Such research focuses on high-mobility and high speed indexes and, for the first time, has been able to realize a jumping-over-water action, as performed by a machine-dolphin. The simulated robot-fish has an important application in the military reconnaissance field, for example, the ability to quietly get close to offshore aircraft carrier formations.

To promote the application and popularization of AI in China, the National Natural Science Foundation of China initiated and hosted the "China Intelligent Vehicle Future Competition". The road conditions in the competition were very close to those of an actual traffic scene. The 7$^{th}$ competition, held in 2015, attracted 12 research institutes, such as the National University of Defense Technology, Tsinghua University and the Military Transportation University. The number of vehicles that joined in the competition was as high as 20. According to the introduction of the person who was responsible for the challenge, Professor Wang Feiyue, China has been the country that manufactures and sells the largest number of cars in the world for the past 5 years. Last year, the number of newly manufactured cars even broke the previous record and reached 20 million, and the vast popularization of "intelligent traffic" technology is expected to quickly open an AI civil market, with a value reaching hundreds of billions of RMB.

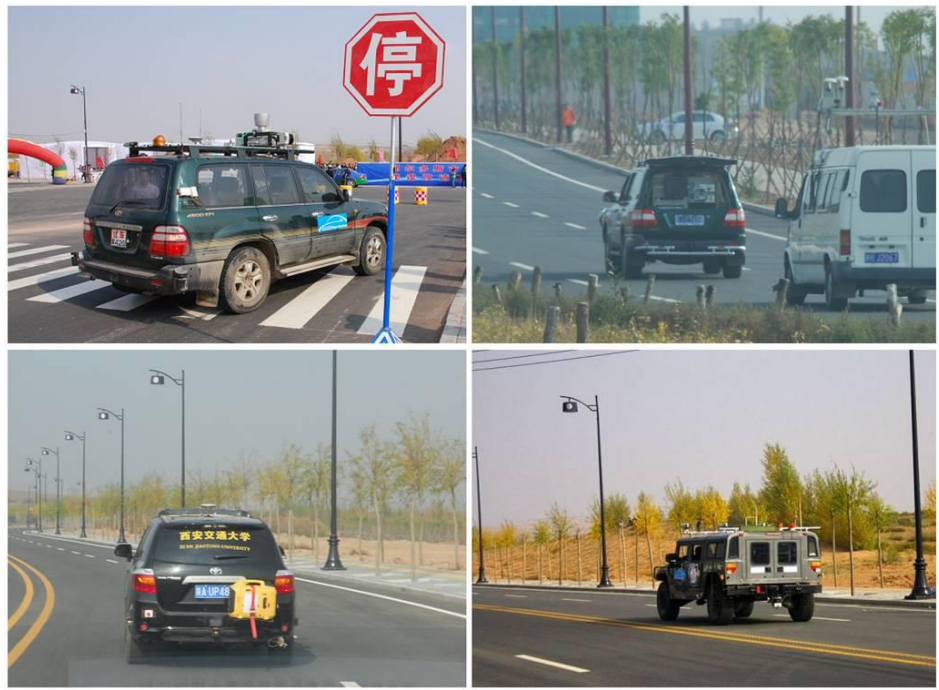

**Fig. 2**. China Intelligent Vehicle Future Competition.

**Inspiration of AI in China**

The AI epidemic has triggered a heated discussion among China's intelligentsia. Wu Gansha, former president of the Intel China Research Center, wrote a paper entitled *Race with the Machine: Win or Die*, sharing his opinion that the intelligence of machines would finally surpass human beings, but he thought that it was important to adopt an attitude of prediction but not forecasting. Yu Kai, former vice-president of the Baidu Research Institute, thought that there must be a clear direction for the development of technology. If humans do not take any control over technology, then AI will turn into an evil thing. However, if humans assume sensible control, then AI will actually represent humanity. Professor and Executive Director from the

School of Software of the Nankai University, Li Qingcheng, who translated the book *The Singularity Is Near: When Humans Transcend Biology* into Chinese, further thought that humanity should guide the development of scientific technology and that the binding of the robot to Asimov's three laws of robotics was not reliable; therefore, it would be necessary to constrain the creators of the robot-humans. At the same time, China's traditional culture stressed the human factors more.

Although current AI has aroused fresh interest around the world, China's industry insiders have a calm understanding. According to the opinion of the president of Microsoft Research Asia, Hong Xiaowen, it was first necessary to make the differences between Intelligence, Intellect and Wisdom clear, while the machine was still at the stage of intelligence. Professor Zhou Zhihua from the Nanjing University simply divided AI into strong AI and weak AI, and holds that the current scientific research is mainly in the area of weak AI. Professor Zhang Changshui from Tsinghua University considers[11] that even for weak AI, current machine-learning methods still faces some challenges, including problems of high-dimensional space, difficulty in seeking the best solution and poor interpretability.

In conclusion, it is very important for us to regard AI as a huge and complicated engineering subject (not simply as a pure scientific subject) formed by many scientific theories, which will have a great influence for AI to be practically possible. From an overall perspective, China is still a follower in the field of AI science. However, from the impressive record of academic publications, state-level investment and the broad participation and support of the whole society, China could catch up quickly in multiple aspects. There are plenty of reasons to believe that, based on centralized support by the national resources in computing ability, big data, labor cost, China will achieve top rankings in AI engineerings soon. China will do a very good job by applying AI in engineerings with the faith of Pragmatic. Moreover, under the influence of the Confucian school of thought's golden mean and harmony over the past thousands of years, human intelligence based on carbon and AI based on silicon could coexist in China, with neither replacing the other. In the future, they will mix and evolve into a new form of intelligence. Such intelligence would not only give play to the powerful ability of memory, computing and recognition but would also develop and evolve toward people and foremost toward humanism.

**Acknowledgments:** This work is supported by National Natural Science Foundation of China (NSFC No. 61272049), and the joint research fund for UCAS and CAS institute (Y55201TY00).

Huai-Yu Wu is an Associate Professor in the National Laboratory of Pattern Recognition, Institute of Automation, Chinese Academy of Sciences. His research interests fall in the fields of pattern recognition and intelligent systems, 3D visual computing, shape perception and analysis.

Feiyue Wang is currently a Professor and Director in the State Key Laboratory of Management and Control for Complex Systems, Institute of Automation, Chinese Academy of Sciences. He is a member of Sigma Xi and an Elected Fellow of IEEE, INCOSE, IFAC, ASME, and AAAS.

Chunhong Pan is a Professor in the National Laboratory of Pattern Recognition, Institute of Automation, Chinese Academy of Sciences. His current research interests include computer vision, image processing, computer graphics, and remote sensing.